# Making Smartphone Application Permissions Meaningful for the Average User

Amer Chamseddine and George Candea
School of Computer and Communication Sciences
École Polytechnique Fédérale de Lausanne (EPFL), Switzerland

## Abstract

Smartphones hold important private information, yet users routinely expose this information to questionable applications written by developers they know nothing about. Users may be tempted to think of smartphones as old-style dumb phones, not as powerful network-connected computers, and this opens a gap between the permissions-based security paradigm (offered by platforms like Android) and what users expect. This makes it easy to fool users into installing applications that steal their information. Not surprisingly, Android is now a more favored target for hackers than Windows [14].

We propose an approach for closing this gap, based on the observation that the current permissions system—rooted in good ol' UNIX-style thinking—is both too coarse and too fine grained, because it uses the wrong axes for defining the permissions space. We argue for replacing the paradigm in which "an app accesses device resources" (which is foreign to most non-geeks) with a paradigm in which "an app accesses user-tangible services." By using a simple piece of middleware, we can wrap this view of application control around today's permission system, and, by doing so, no conceptual refactoring of applications is required.

## 1 Introduction

Mobile applications ("apps") have captured the mind and soul of consumers, with much of the competition in the smartphone arena revolving around the variety and coolness of apps available on a given device. An average smartphone owner has more than 40 applications installed [1], actively uses 15 of them, and spends more than ten hours a month interacting with these applications—more time is spent with apps than spent talking on the phone or using it to browse the Web [13].

Most users do not know what exactly their installed apps do or have access to. Do they read the address book and send the contacts to spammers? Do they track the user's movements via GPS? Do they turn on the phone's camera/microphone and spy on the user? This is serious, because smartphones, unlike computers, are always on, network-connected, and always with their user.

Today's smartphones are powerful computers, but it is easy for users to think of them as "dumb" phones; the magic by which apps can dynamically augment the functionality of this "phone" is not fully understood and is therefore ignored. Certain concepts, such as "app permissions" do not really have an equivalent in the world of traditional phones. Discrepancies of this sort create a gap that, from a security standpoint, invites exploitation. We aim to bridge this gap by making the concept of app and app permissions more "natural" to users, thus making it easier to understand and manage.

We start by looking at Android, both because of its popularity and because we think it has important weaknesses in this context. Android permissions were conceived to mitigate the threats described above, by putting the user in control: users must explicitly grant the app, at installation time, access to the resources it requires. Unfortunately, many Android permissions are too coarse grained, thus leading to a violation of the "least privilege" principle. For example, a weather forecast application may simply need to download data from a specific server, but in the current Android Permissions System a user would have to give this app full Internet access. Other permissions are too technical for average users, like "use SIP service" or "change the Z-order of tasks." A puzzled user trying to choose between saying "no" and getting to (install and) use the app will likely opt for the latter. And, in so doing, the user may unwittingly hand over to the app control over his/her data.

Our goal is to eliminate the mismatch between user expectations and actual app behavior, as pertains to user data. We describe a split/merge ap-proach for morphing the Android Permissions System into user-meaningful permissions: *split proxies* on the phone turn permissions that are too coarse into finer ones, and *merge proxies* combine low-level or too-technical permissions into semantically meaningful ones. Permissions and proxies form a hierarchy, and applications access the top layer of this hierarchy. We envision service providers (such as AdMob, Facebook, or Google Analytics) defining the permissions by writing such proxies and providing them to developers as part of their SDKs.

Before describing our approach in more detail, we briefly present the Android Permissions System.

## 2 Android Permissions: The Good and the Bad

Each mainstream smartphone platform has its own security model, with pros and cons. iOS and Windows Phone 8 control apps' behavior largely by testing them before admitting them in the respective app store/marketplace; users trust Apple and Microsoft to do proper checking of the apps. Of course, full verification of an app is still an open challenge, and apps can still misbehave [15]. Android, however, defers the choice for what an app is allowed to do to the end user. We believe the ideal smartphone security model will offer users some level of choice, and thus the question of how to formulate and grant permissions will remain pertinent. This makes Android an interesting research target.

### 2.1 Android Permissions System: An Overview

The Android Permissions System [2] allows developers to specify a list of accesses and permissions their app needs in order to function properly; this list is part of the app's manifest file. These permissions control access to sensitive device APIs, such as the camera, the microphone, or GPS sensor.

Which permissions are granted to an app is decided upfront, at app installation time—after installation, apps cannot request additional permissions. When a user installs an app, the operating system displays the list of requested permissions (see for example Figure 1). Presumably, the user reads the list, makes an informed decision on whether to grant or not the requested permissions, and clicks Install or Cancel. Installation can only proceed if the user accepts all requested permissions.

Underneath the covers, Android (which runs a modified Linux kernel) enforces its security policies by mapping applications and permissions onto UNIX-style users and groups, respectively. When installing a new app, Android creates a new userid for it, and runs the app as that user. Android maintains one usergroup for each permission and, if the user grants a particular permission at install time, the app's userid is added to the respective usergroup.

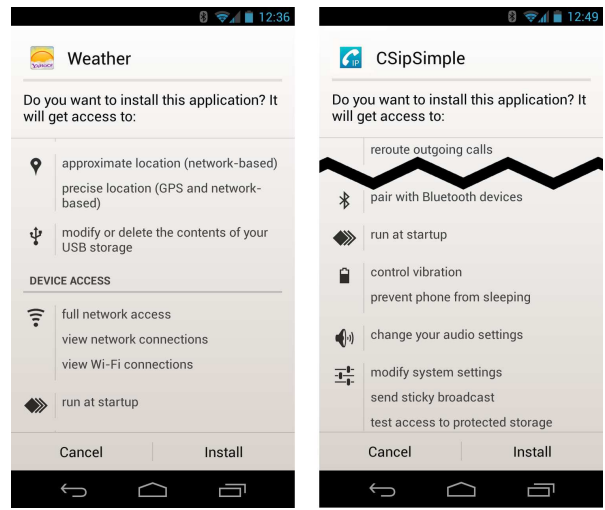

**Figure 1:** Installing Android apps: On the left, a weather forecast app requests full access to the Internet (unnecessarily coarse grained). On the right, a VoIP app requests access to "send sticky broadcast," "reroute outgoing calls," etc. (too technical for an average user).

When the app invokes system services, the kernel checks whether the app is entitled to make the respective system call or not (e.g., if an app tries to create a network socket but its userid does not belong to the "Internet access" usergroup, Android denies the socket creation attempt). Enforcement is performed in the kernel so that even native code cannot bypass the permissions system.

The benefit of the Android Permissions System is that the operating system promises users that each app will only be allowed to access the APIs that were granted by the user at install time. Furthermore, a user can determine at any time what permissions an app has by looking at its list of permissions, which is available post-installation.

### 2.2 Shortcomings

Unfortunately, most users are not prepared to exercise their right to choose, because some Android permissions can be incomprehensible to them. These typically relate to device details that are outside the user's realm of comprehension (access SurfaceFlinger, broadcast WAP PUSH receipt notifications, perform I/O over NFC, etc.). In a recent study, only 17% of users were aware of the requested permissions list and, of those, only 3% had some understanding of the permissions' meaning, power, and consequences [9]. In our experience, even technical users make the decision based on intuition rather than careful analysis. This is because, even when understanding what the permissions are about, reasoning about how a combination of permissions might compromise privacy or security is difficult.

At the same time, some Android permissions force the violation of the "least privilege" principle by being too coarse-grained, as illustrated by the weather forecast app in Figure 1. Another common case is when a non-networked app uses a library/SDK to show ads (e.g., via AdMob) or to collect usage statistics (e.g., via Google Analytics): even though the user knows the app should not require Internet access, the fact that it is ad-supported forces it to ask for such access. This way, users are trained to accept that any reasonable app needs Internet access. This is problematic, given that almost 25% of today's free Android apps request access to private user information (location data, demographic information, etc.) claiming it is used for targeted ads only, but this data ends up being sent to the app provider's servers [11, 5, 4]. Careful users have little choice too: by being conservative and rejecting such apps, one is deprived by much of what makes a smartphone compelling.

The current Android Permissions System is *both too fine-grained and too coarse-grained* at the same time. This paradoxical situation stems from the permissions being defined along the (in our opinion) wrong axes: the permissions system is thought in terms of apps accessing device resources, and this is a highly OS-centric view of what the smartphone does. It leaves to the user to extrapolate which user-visible services could be accessed via those resources, or which user data could be read/written. We argue that, instead, the permissions system should use as primitives those abstractions and services that are user-visible, such as ad services, web sites that can be browsed by the user, etc.

## 3 Our Approach: Split / Merge Proxies

Our goal is to enable users to more easily map permissions to "tangible" services. To make this possible, our system is a new layer introduced between the OS and the apps (Figure 2). It consists of a hierarchy of permission proxies. A permission proxy takes permissions from a lower layer and transforms them into a new permission exposed to the upper layer. We have two types of permission proxies: *split proxies* that divide permissions into finer-grain ones, and *merge proxies* that combine multiple permissions into a semantically higher-level one.

### 3.1 Proxy-based Permission System

Proxies are built as thin Android apps that leverage Android's flexibility for defining new permissions, thus not requiring changes to the OS in most cases. Each proxy $X$ declaring a new permission $P$ also defines an API corresponding to the use of $P$. For example, as will be detailed later, using a selective form of HTTP access would require using the `SelectiveHttpClient` class provided by the corresponding proxy, instead of the `DefaultHttpClient` class. When an app requests permission $P$, our system ensures that the app uses the API exposed by $P$, and prevents it from circumventing $X$ to go straight to the APIs used by $X$. In other words, $X$ interposes hermetically between the app and the permission(s) underneath $X$.

Apps are generally written using various SDKs (for example, the Google Analytics SDK), and we envision such SDKs defining the proxy/proxies they need. The proxies themselves will typically be written by the provider of the SDK, but it is possible for SDKs from different providers to use the same proxy/proxies. For example, if Android were to provide the Domain-Selective Internet Access permission (described below), advertising companies could build their SDKs using this permission. An app then uses the SDK as a library and may well be oblivious to the existence of the proxy, because the SDK itself invokes the proxy-specific API.

To make proxy distribution efficient, we envision them being in a special "Proxies" section of the Play Store, and SDKs ship with metadata describing which proxies they require. When installing an app using that SDK, if the required proxy(ies) is not already on the device, it gets retrieved and installed. When a new proxy is installed, it requests permissions just like Android apps do today, and the user can choose to grant them or not. In other words, proxies get no special privileges. Split proxies, such as the Domain-Selective Internet Access proxy, could as well be incorporated in Android directly.

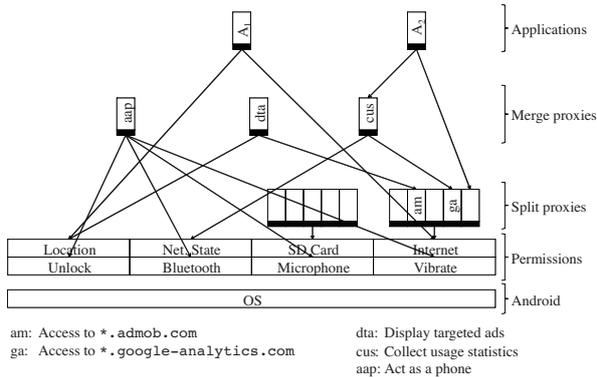

**Figure 2:** Our proposed Split/Merge Permissions Proxy architecture. Here, $A_1$ follows the traditional style of Android permissions, and requests its permissions directly from the OS, whereas $A_2$ requests the semantically meaningful permissions to collect usage statistics and communicate with its server.

Compared to today's permission system, our proposed system requires strictly less trust from the end user, because no new low-level permissions are created. We expect there to be much fewer proxies than apps (proxies can be thought of as libraries) and would be written and signed by well-known, trustworthy companies. Even when proxies are written by untrusted developers, they are still small pieces of code (a few hundred LOC), have no UI, and can be easily verified using code inspection and/or specialized tools [8, 7], and then certified by the app store provider or the handset manufacturer; some of them could ship with Android.

### 3.2 Split Proxies

If a permission is too coarse to enable the "least privilege" principle, it needs to be split. We illustrate here how we split the general Internet access permission using the Domain-Selective Internet Access proxy, and split storage access using the Selective SD Card Access proxy.

Android offers one permission for all network communication with the outside world; an app with this permission can perform any network activity, ranging from GET-ing data over HTTP to listening on server sockets. We use the ***Domain-Selective Internet Access*** proxy to split this permission into finer-grained ones, based on the domains that the app wants to communicate with. The proxy accepts wildcard specifications, to enable communication with many subdomains. For example, a BBC news app may request, at installation time, network access to "`*.bbc.co.uk`."

The situation is similar when using phone storage, and apps use the SD card in a rather chaotic manner. Many applications create a folder on the root of the SD card and use it to store files, configuration files, or even backup files, with no naming convention at all. Even though newer versions of Android specify standard ways of storing files on the SD card, most applications on the Play Store still use their own conventions. The SD card can contain highly sensitive data, such as pictures, videos, recorded conversations, and downloaded files. Just because an application needs to store files on the SD card does not mean that it should have full access to the user's personal files or to another application's files stored on the SD card. We use the ***Selective SD Card Access*** proxy to split the SD card permission into finer-grained, folder-level access permissions: each app gets automatically assigned its own folder, named `/sdcard/Android/data/<app_pkg>/files/` in accordance with the Android convention, and all the app's storage I/O is restricted to this folder and its subfolders—in essence `chroot`-ing the app.

### 3.3 Merge Proxies

A merge proxy takes permissions that are too low-level/technical for average users and combines them into permissions that are more meaningful. We illustrate this with two examples, the Collect-Usage-Statistics and Act-as-a-Phone proxies.

An app developer may want to use Google Analytics to gather information on how people use him/her app, which features are used most often, etc. Since this data is reported to Google over the network, it means that all such apps must request up to two permissions in addition to what they normally need: Full Internet Access and Network State Access (to determine when buffered statistics can be flushed). The ***Collect-Usage-Statistics*** proxy, however, layers on top of the Domain-Selective Internet Access proxy (with access to `*.google-analytics.com`) and the native Network State Access permission. It defines a new permission, called Collect-Usage-Statistics. Apps using Google Analytics can now, upon installation, request permission to "collect usage statistics" instead of "full Internet access" and "network state access."

As another example, VoIP applications usually request a lot of permissions in order to simulate a phone call: They need the permission to unlock the screen and ring the device when a phone call arrives. Once the call is connected, they need the microphone recording permission (to capture the user's voice) and the Bluetooth permission (if a wireless headset is used). We introduce the *Act-as-a-Phone* merge proxy to combine all these permissions. VoIP applications (such as CSipSimple, Skype, and others) can now request the Act-as-a-Phone permission, instead of requesting a long list of highly technical permissions. It is also conceivable for a merge proxy to restrict certain aspects of the underlying permissons, such as only allowing an app that is "acting as a phone" to turn on the microphone in response to an explicit user action.

### 3.4 User-Controlled Permission Slicing

In addition to provided proxies, our permission system is also amenable to (expert) users deciding on the fly to only selectively grant requested permissions. For example, a legacy app that does not use our proposed system may request full Internet access, but the user decides to give it access only to specific domains that are deemed safe. If the application never attempts to access any other domains, all is OK, and the user is safe. If the application attempts to access other domains, then blocking those accesses would likely prevent it from functioning—instead, our system can prompt the user, much the way a personal firewall would do, and (upon the user's request) block the connection outright, let it go through, or fake the connection and pretend that the host is unreachable, thereby giving the app a chance to recover and continue operating.

### 4 Prototype and Early Results

As a first step, we implemented the *Domain-Selective Internet Access* split proxy and the *Collect-Usage-Statistics* merge proxy.

Many Android apps, when communicating with their server side, employ the `DefaultHttpClient` class. We built Domain-Selective Internet Access as described earlier, and it provides its own implementation of the `HttpClient` interface, `SelectiveHttpClient`, which extends the same classes as `DefaultHttpClient` and offers the same functionality. This class communicates directly with our Domain-Selective Internet Access proxy. The proxy requests Full Internet Access from the OS and exposes the Domain-Selective Internet Access permission. The proxy is essentially an Android service that other applications can bind to. The service interface is defined using the Android Interface Definition Language.

Apps using `SelectiveHttpClient`, instead of requesting Full Internet Access, only request the Domain-Selective Internet Access permission. `SelectiveHttpClient` routes all traffic through the proxy, which checks that the access requests are being made to the domains approved by the user; if yes, the requests are forwarded to the network.

The Collect-Usage-Statistics proxy is also built as a normal Android app that requests a Domain-Selective Internet Access permission and the Network State Access permission, and exposes a new Collect-Usage-Statistics permission. In our preliminary prototype, the Collect-Usage-Statistics proxy only reports to Google Analytics. The communication between apps and this proxy is done through Android's standard Intent Messaging API.

We analyzed many real-world apps (like Facebook, Skype, WhatsApp, etc.) and found that they all request Full Internet Access permission. 73% of the Play Store apps are free, and, of those, 80% are ad-supported [12]. These applications end up gaining access to sensitive personal data, such as contacts and email, GPS location, demographic data, etc, without legitimately needing it [10]. Having the Domain-Selective Internet Access permission makes it possible to guard users' information, without depriving users of the wide variety of free applications or preventing developers from generating revenue by displaying ads.

There is one caveat to the Domain-Selective Internet Access permission: since Android allows apps to modify the device's DNS server when connected to a WiFi network, it becomes possible for a malicious app to subvert the proxy: if it can enlist the cooperation of a rogue DNS server that supplies a fake IP address when queried for a domain, the app could access servers outside the list that the user approved. In other words, this proxy is as secure as DNS itself. This weakness can be fixed by using DNSSEC or disabling the ability of apps to change the DNS server configured on the phone.

**Performance.** From a performance perspective, going through a proxy normally introduces overhead. However, since most of the proxied operations are bottlenecked by the latency of I/O anyway, the measured overhead introduced by our proxies is not perceivable. To time HTTP inter-actions with and without the proxy, we created 32 `DefaultHttpClient` objects and used them to issue 32 POST requests; the average time of the entire interaction (all 32 POSTs) was on the order of 30 ms. In the case of the Domain-Selective Internet Access, most of the functionally essential operations deal with sockets, and have essentially the same latency and throughput, with or without the proxy. For Selective SD Card Access, the bottleneck is in the file I/O operations. Since smartphone applications usually have a GUI, they spend most of their time waiting for user interaction; we expect that, even if future proxies introduce some overhead, it would not be user-perceptible.

**Deployment.** We believe our proposed system is easy to adopt and deploy today. Writing the proxies is relatively straightforward and requires little code (the Domain-Selective Internet Access proxy has less than 500 LOC). Non-malicious apps should not need to be conceptually refactored in order to use the proposed permission system; at most they may need to replace the use of some interfaces with those provided by the proxies. From the developers' perspective, these changes are worthwhile because they make apps more trustworthy. Unlike today (when users are "trained" to trust questionable apps), our proposed system may gradually push users to distrust apps that bypass the proxies and request direct access to low-level resources.

**Manageability.** Permission proxies are either part of Android or part of SDKs. We can also imagine an after-market for "better" versions of the same proxies; users can then substitute one proxy for another. We estimate that a "power user" of Android smartphones will require about a couple dozen permission proxies, which compares well to the set of 130 permissions currently defined by Android [3].

**Permissions Calculus.** As mentioned earlier, it is difficult for users to reason in their head (even when they understand what a set of permissions is about) about how a combination of permissions might compromise privacy or security. Our system outsources this task to proxy developers who must now think about whether combining permissions does not open up holes. It is of course good to have this done by trusted proxy developers, but we also believe there is an opportunity to employ a formal approach—we are currently looking into using a variant of BAN logic [6] for this.

## 5 Conclusion

We presented an approach for turning hard-to-understand low-level permissions of smartphone apps into ones that are semantically meaningful to average users. We propose a layer of proxies between the OS and apps, which can then merge and split low-level permissions in a way that matches user expectations. The proxies are small, verifiable, and require no conceptual refactoring of apps.

**References**

[1] J. Aimonetti. Nielsen: 1 in 2 own a smartphone, average 41 apps. *CNET Technology Reviews*, May 2012. http://reviews.cnet.com/8301-19512_7-57435397-233/nielsen-1-in-2-own-a-smartphone-average-41-apps/.

[2] Android permissions. http://developer.android.com/guide/topics/security/permissions.html.

[3] Android permissions reference. http://developer.android.com/reference/android/Manifest.permission.html.

[4] Bit9. Pausing Google Play: More than 100,000 Android apps may pose security risks. https://www.bit9.com/files/1/Pausing-Google-Play-October2012.pdf, Oct. 2012.

[5] T. Bradley. Study finds 25 percent of android apps to be a security risk. *PCWorld*, Nov. 2012. http://www.pcworld.com/article/2013524/study-finds-25-percent-of-android-apps-to-be-a-security-risk.html.

[6] M. Burrows, M. Abadi, and R. Needham. A logic of authentication. *ACM Transactions on Computer Systems*, 8(1):18–36, Feb 1990.

[7] C. Cadar, D. Dunbar, and D. R. Engler. KLEE: Unassisted and automatic generation of high-coverage tests for complex systems programs. In *OSDI*, 2008.

[8] V. Chipounov, V. Georgescu, C. Zamfir, and G. Candea. Selective symbolic execution. In *HotDep*, 2009.

[9] A. P. Felt, E. Ha, S. Egelman, A. Haney, E. Chin, and D. Wagner. Android permissions: User attention, comprehension, and behavior. In *SOUPS*, 2012.

[10] K. J. Higgins. http://www.darkreading.com/mobile-security/167901113/security/privacy/240012705/more-than-25-of-android-apps-know-too-much-about-you.html, Nov. 2012.

[11] B. Levine. Be wary of free Android apps. http://www.sci-tech-today.com/story.xhtml?story_id=13200006Q1XO.

[12] I. Lunden. In mobile apps, free ain't free, but cambridge university has a plan to fix it. *TechCrunch*, Mar. 2012. http://techcrunch.com/2012/03/06/in-mobile-apps-free-aint-free-but-cambridge-university-has-a-plan-to-fix-it/.

[13] A. Mindlin. Using phones, but not to talk or surf. *The New York Times*, Mar. 2011. http://www.nytimes.com/2011/03/07/business/media/07drill.html.

[14] D. Reisinger. Android overtakes Windows as top threat. MIT Technology Review, Dec. 2012. http://www.technologyreview.com/view/508316/the-changing-face-of-security-android-overtakes-windows-as-top-threat/.

[15] R. Ritchie. iOS 6 wants: Granular privacy control. *iMore*, Feb. 2012. http://www.imore.com/path-apps-accessing-contacts-inspiration-android.